\documentclass[12pt]{iopart}

\usepackage{graphicx}
\usepackage{dcolumn}
\usepackage{bm}
\usepackage[utf8]{inputenc}
\usepackage[T1]{fontenc}
\usepackage{amsthm,amsmath,amssymb,amsfonts}
\usepackage{mathrsfs}
\usepackage{cite}
\usepackage{color}
\usepackage{float}
\large
\usepackage{multirow}
\usepackage{booktabs}

\begin{document}

\title[Impurity turbulent transport and its role in driving zonal flows in tokamak plasmas]{Impurity turbulent transport and its role in driving zonal flows in tokamak plasmas}

\author{Yuesong Li$^{1,2,\dag}$, Zihao Wang$^{1,2,\dag}$, Haochuan Wang$^{1,2}$, Debing Zhang$^3$ and Shaojie Wang$^{1,2,*}$}
\address{$^1$Department of Engineering and Applied Physics, University of Science and Technology of China, Hefei 230026, China.}
\address{$^2$National Key Laboratory of Frontier Physics in Controlled Nuclear Fusion, Chinese Academy of Sciences, Hefei 230031, China.}
\address{$^3$School of Physics, East China University of Science and Technology, Shanghai 200237, China.}
\address{$\dag$These authors contribute equally to this paper}
\address{$^*$Author to whom any correspondence should be addressed.}
\ead{wangsj@ustc.edu.cn}
\vspace{10pt}
\begin{indented}
\item[]November 2026
\end{indented}

\begin{abstract}
The driving mechanisms of zonal flows (ZFs) in multi-ion-species tokamak plasmas are investigated by using gyrokinetic theory and simulations. For deuterium plasmas containing either helium ash with a steep density gradient or lithium impurity at a high charge concentration, the magnitude of impurity turbulent energy flux (TEF) can be comparable to, or even stronger than, that of the main-ion TEF. Moreover, the impurity TEF significantly drives ZFs. Both the impurity and main-ion turbulent poloidal Reynolds stress (PRS) drives are significant for the case with a high lithium charge concentration. The shielding timescale of the turbulent PRS drive is found to be the damping time of  geodesic acoustic mode, which is significantly shorter than one ion bounce period.
\end{abstract}

%
%
%
\maketitle
%
%

\section{INTRODUCTION}\label{sec:introduction}
Anomalous transport driven by drift-wave turbulence \cite{Horton1999RMP,Liewer1985NF}, e.g., ion-temperature-gradient (ITG) turbulence, is a key obstacle to energy confinement in tokamak plasmas. Zonal flows (ZFs), with an axisymmetric ($m = n = 0$, where $m$ and $n$ are the poloidal and toroidal mode numbers, respectively) and radially varying structure, are widely regarded as a self-regulation mechanism of turbulence \cite{Diamond2005PPCF,Lin1998Science} and are therefore important for improving plasma confinement, e.g., the L-H transition \cite{Wagner1982PRL} and the formation of internal transport barriers \cite{Levinton1995PRL,Burrell1998PPCF,Takenaga2009NF,Han2022nature,Mazzi2022NP}. ZFs are linearly stable but can be nonlinearly driven by ITG turbulence. ZFs can be excited by modulational instability \cite{Chen2000POP} and eigenmode self-interaction \cite{Ajay2020JPP,Wang2022PRE}, and, on the mesoscale, nonlinearly driven by the turbulent poloidal/toroidal Reynolds stress (PRS/TRS) \cite{Diamond1994PRL,Wang2017POP,Zhang2020NF} and turbulent energy flux (TEF) \cite{Wang2017POP}. It is found that the driving effect of the turbulent PRS is shielded by the neoclassical polarization (NP) effects in toroidal geometry \cite{Itoh1996PPCF}. The NP factor given by Rosenbluth-Hinton \cite{RH1998PRL} is $\epsilon_r =1+1.6q^2/\sqrt{\epsilon}$, with $q$ the safety factor and $\epsilon=r/R$ the inverse aspect ratio. Further gyrokinetic simulations indicate that the nonlinear driving effect of the turbulent PRS in toroidal geometry is timescale-dependent \cite{Wang2025NF}. On the timescale longer than the ion bounce period, the turbulent PRS is shielded, whereas on shorter timescales, it is not. Unlike the turbulent PRS, the driving effects of the turbulent TRS and TEF are not shielded by NP effects \cite{Wang2017POP}.

All the studies above were carried out in pure plasmas. In fact, impurities are ubiquitous in tokamak plasmas, mainly arising from plasma-wall interaction \cite{Mazzitelli2011NF} and fusion products \cite{Synakowski1995PRL,Budny2008NF}. Tokamak experimental observations \cite{Murakami2001NF,Chen2024NF} have shown that the presence of impurities has significant effects on ZFs. In the DIII-D tokamak \cite{Murakami2001NF}, the presence of impurities enhances the shearing rate of ZFs. In the HL-2A tokamak \cite{Chen2024NF}, the amplitude of the low-frequency ZF-like structure dramatically increases in the presence of impurities, which is attributed to the enhanced nonlinear energy transfer from turbulence via the turbulent PRS in the presence of impurities. Moreover, several simulations have investigated the effects of impurity charge concentration and impurity density gradient on ITG turbulence \cite{Romanelli2011NF,Romanelli2013NF,Bonanomi2018NF}, as well as the effects of impurities on ZFs \cite{Kim2017POP,Seo2022POP}. It is found that the presence of impurities makes the ZF staircase structure more stable but reduces the ZF amplitude \cite{Kim2017POP,Seo2022POP}. However, these studies have not clarified how impurities affect the driving effects of turbulent transport fluxes, particularly the TEF, on ZFs. To further elucidate the impurity-induced changes in ZFs found in the experiment \cite{Chen2024NF} and simulations \cite{Kim2017POP,Seo2022POP}, it is necessary to investigate how impurities affect the ZF driving.

Recently, several gyrokinetic theories have investigated the effects of impurities on ZFs driven by the initial ion density \cite{Guo2017NF,Li2026POP} and temperature perturbations \cite{Li2026POP}. The results show that the presence of impurities enhances the plasma classical polarization and thus reduces the driving effects of the main ion perturbations on ZFs \cite{Li2026POP}. Note that the model driving sources in \cite{Li2026POP}---initial density and temperature perturbations representing the turbulent PRS and TEF, respectively---were artificially prescribed. In fact, the turbulent transport fluxes (including the turbulent PRS, TRS, and TEF) and ZFs evolve self-consistently. Therefore, the effects of impurities on the ZF driving should be investigated in a self-consistent turbulence field.

In this work, we have investigated the driving mechanisms of ZFs during the self-consistent relaxation of ITG turbulence in the presence of impurities for the first time. Gyrokinetic theory shows that both the impurity and main-ion turbulent transport fluxes drive ZFs significantly in certain cases. On this basis, two physically distinct impurity cases are considered in the subsequent simulations: (i) fully stripped lithium (Li$^{3+}$) at a high charge concentration ($Z_{\rm z}n_{\rm z}/n_{\rm e}=50\%$), motivated by the initial phase of an FTU discharge with the liquid lithium limiter \cite{Mazzitelli2011NF}; (ii) low-concentration helium ash (He$^{2+}$) with a steep density gradient, representing a scenario of fusion-product accumulation \cite{Budny2008NF}.  The simulated ZF structure and amplitude agree well with the theoretical predictions. The main findings are as follows. First, in both impurity scenarios, the impurity particle flux is outward, whereas the main-ion particle flux is inward, consistent with the quasi-neutrality condition. These oppositely directed particle fluxes contribute to an impurity TEF whose magnitude is comparable to or even larger than the main-ion TEF. Moreover, the impurity TEF drives ZFs significantly in the quasilinear and saturation stages of ITG turbulence. Second, both the impurity and main-ion PRS drives are significant in the nonlinear stage for the case with a high lithium charge concentration, and both of them are time-dependently shielded by the NP effects. The shielding timescale of the turbulent PRS drive is found to be the damping time of geodesic acoustic mode (GAM), which is significantly shorter than one ion bounce period.

The remainder of this paper is organized as follows. In section \ref{sec:theory}, the derivation of the ZF driving equation in multi-ion-species tokamak plasmas is presented. Subsequently, the driving effects of the impurity turbulent transport fluxes, as well as the time-dependent shielding of the turbulent PRS drive, are discussed. In section \ref{sec:simulation}, the simulation code NLT is briefly introduced, and the nonlinear simulation results of TEF evolution and the ZF driving in two impurity scenarios are presented. Section \ref{sec:summary} presents the summary and discussion.

\section{The zonal flow driving equation in multi-ion-species tokamak plasmas}\label{sec:theory}
In this section, we derive the ZF driving equation in multi-ion-species tokamak plasmas by using gyrokinetic theory. The derivation proceeds in three steps. We first solve the gyrokinetic Vlasov equation for the zonal distribution function, and express its turbulent driving source through the moments of the turbulent fluctuating distribution function. Then we couple the different particle species by the quasi-neutrality equation. Finally, by solving for the zonal potential, we obtain the temporal asymptotic ZF and its time-dependent generalization.

Following \cite{RH1998PRL}, we begin with the ensemble-averaged Frieman–Chen gyrokinetic equation \cite{Frieman1982POF} for the zonal part of the perturbed distribution function
\begin{equation}\label{eq:GKE}
\partial_tg_{j}+ \mathcal{L} _{0,j}g_{j}=\frac{e_j}{T_{j}}F_{0,j}J_{0}\partial_t\delta\phi+J_{0}S_{j}F_{0,j},
\end{equation}
with $j={\rm i,z,e}$ denoting the main ions, impurities and electrons, respectively. $\mathcal {L}_{0,j}\equiv(\textit{v}_{\parallel}\boldsymbol b+\bm{V} _{{\rm d},j})\cdot \nabla$, with $v_{\parallel}$ being the parallel velocity and $\bm{b}=\bm{B}/B$. $\boldsymbol B=\boldsymbol B_T+\boldsymbol B_P$ is the equilibrium magnetic field, with $\boldsymbol B_T$ and $\boldsymbol B_P$ denoting toroidal and poloidal magnetic field, respectively. $\boldsymbol V_{{\rm d},j}$ is the magnetic drift velocity in the equilibrium magnetic field. The perturbed distribution function for ZFs is $\delta f_{j}=-(e_j\delta \phi/T_j)F_{0,j}+g_{j}(\boldsymbol X,K,\mu,t)$. The equilibrium distribution function is taken as the local Maxwellian $F_{0,j}=\pi^{(-3/2)}v_{{\rm th},j}^{-3}n_j\exp(-K/T_{j})$, with $m_jv_{{\rm th},j}^2/2=T_j$. $n_{j}$ and $T_{j}$ are the equilibrium density and temperature, respectively. $m_j$ is the particle mass. $K=m_jv_{\parallel}^2/2+\mu B$, with $\mu$ being the magnetic moment. $\delta \phi$ is the zonal perturbed electrostatic potential, which is assumed to be given by the eikonal form $\delta \phi(\boldsymbol x)=\delta \phi _k\exp({\rm i}k_rr)$. $\boldsymbol X=\boldsymbol x-\boldsymbol \rho_{{\scriptscriptstyle \rm L},j}$ is the guiding center position, with $\boldsymbol x$ being the particle position and $\boldsymbol \rho_{{\scriptscriptstyle \rm L},j}=\bm{b}\times m_j\bm{v}/e_jB$. $\bm{v}$ and $e_j$ are the particle velocity and charge, respectively. $k_r$ is the radial wavenumber. $J_{0}=J_{0}(k_r\rho_{{\scriptscriptstyle \rm L},j})$ is the zeroth-order Bessel function, which represents the gyroaverage.

\indent The source term $S_j$ in equation \eqref{eq:GKE}, which represents the driving effects of ITG turbulence on ZFs, is given by \cite{Wang2017POP,Zhang2020NF}
\begin{equation}\label{eq:source}
\begin{split}  
    S_j = 
    &-\frac{1}{r} \partial_r \left\{ r \left[ \frac{2}{3} Q_{r,j} + \left( \frac{K}{T_j} - \frac{5}{2} \right) \left( \frac{2}{3} Q_{r,j} - \Gamma_{r,j} T_j \right) + v_\parallel \Pi_{r\zeta,j} \right] \right\} \frac{1}{p_j}\\
    &+ \frac{1}{e_jBn_j} \frac{1}{r}\partial_r \left[r \frac{1}{r} \partial_r (r \Pi_{r\theta,j}) \right],
\end{split}
\end{equation}
with $p_j$ denoting equilibrium pressure.  $\Gamma_{r,j},\Pi_{r\theta,j},\Pi_{r\zeta,j}$ and $Q_{r,j}$ denote turbulent particle flux, turbulent PRS, turbulent TRS, and TEF, respectively. They are defined as follows

\begin{equation}
\Gamma_{r,j} = \int \mathrm{d}^3 \boldsymbol v \, \langle \tilde {V}_r \tilde{f_j} \rangle_{\mathrm{en}},
\end{equation}

\begin{equation}\label{eq:PRS}
    \Pi_{r\theta,j}=\int {\rm d}^3\boldsymbol v F_{0,j} m_j\left\langle\tilde {V}_r\tilde {V}_{\theta}\right\rangle_{{\rm en}},
\end{equation}

\begin{equation}\label{eq:TRS}
\Pi_{r\zeta,j} = \int \mathrm{d}^3 \boldsymbol v \, \langle \tilde {V}_r \tilde{f_j} \rangle_{\mathrm{en}} m_j\, v_\parallel \frac{B_T}{B},
\end{equation}

\begin{equation}
Q_{r,j} = \int \mathrm{d}^3 \boldsymbol v \, \langle \tilde {V}_r \tilde{f_j} \rangle_{\mathrm{en}} \, K,
\end{equation}
with $\tilde {V}_{r}$,$\tilde {V}_{\theta}$ representing the radial and poloidal components of the turbulent fluctuating $\bm{E}\times\bm{B}$ velocity, respectively. $\tilde {V}_{s}=-\nabla(J_0\delta\phi)\times \boldsymbol B/B^2\cdot \nabla s$ with $s=r,\theta$. $\tilde f_j$ is the turbulent fluctuating distribution function. $\left\langle\right\rangle_{\rm en}$ denotes the ensemble average. Note that the turbulent particle flux $\Gamma_{r,j}$ satisfies the ambipolar condition $\sum_{j=\rm i,z,e}e_j\Gamma_{r,j}=0$, whereas $\Pi_{r\theta,j}$ will contribute a non-ambipolar turbulent particle transport \cite{Zhang2020NF}. Equation \eqref{eq:GKE} can be solved by assuming $\partial_t\ll \mathcal {L}_{0,j}\sim 1/\tau_{{\rm b},j}$, with $\tau_{{\rm b},j}=2\pi qR/(\sqrt{\epsilon}C_{{\rm s},j})$ being the bounce period for trapped particles and $C_{{\rm s},j}=\sqrt{T_j/m_j}$. The operator $\mathcal{L} _{0,j}$ can be eliminated by taking an orbital average. The detailed solution procedure is presented in \ref{appen:theory}.

\indent The zonal potential $\delta\phi$ is determined by the magnetic-flux-surface averaged quasi-neutrality equation including main ions, impurities and electrons,
\begin{equation}\label{eq:QNE}
    \sum_{j={\rm i,z,e}}e_j\left\langle -\frac{e_j}{T_j}n_j\delta\phi+\int {\rm d}^3\boldsymbol vJ_{0}(k_r\rho_{{\scriptscriptstyle \rm L},j})g_{j}\right\rangle_{{\rm FA}}=0.
\end{equation}
Here, $\langle \rangle_{{\rm FA}}$ denotes the magnetic-flux-surface average. For mesoscale ZFs, equation \eqref{eq:QNE} is solved by retaining the ion finite-Larmor-radius and ion finite-orbit-width effects to second order $O(k_r^2\rho_{{\scriptscriptstyle \rm L},j}^2)$ and $O(k_r^2\Delta r_j^2)$, respectively, where $\Delta r_j$ denotes the orbit width. For electrons, we set $k_r\rho_{{\scriptscriptstyle \rm L},\rm e}=0$ and $k_r\Delta r_{\rm e}=0$, neglecting their contributions to the polarization. The detailed solution procedure of equation \eqref{eq:QNE} is similar to that in \cite{Wang2017POP} and \cite{Zhang2020NF}, to which readers are referred for details. The zeroth-order terms in equation \eqref{eq:QNE} yield $\sum_{j={\rm i,z,e}}e_j r^{-1}\partial_r(r\Gamma_{{\rm r},j})=0$, the multi-ion counterpart of equation (95) in \cite{Zhang2020NF}. The second-order terms then give the ZF driving equation,
\begin{equation}\label{eq:Er}
\begin{split}
\partial_t \delta E_r = 
&- \frac{1}{\rho_{\scriptscriptstyle \rm M}} \sum_{j=\rm i,z} \frac{m_j}{e_j} \partial_r \left[ \frac{1}{r} \partial_r \left( r \frac{2}{3} Q_{r,j} \right) \right]\\
&+ \frac{1}{\epsilon_r} \frac{B_T}{\rho_{\scriptscriptstyle \rm M}} \sum_{j=\rm i,z} \frac{1}{r} \partial_r (r \Pi_{r\theta,j}) 
- \frac{B_P}{\rho_{\scriptscriptstyle \rm M}} \sum_{j=\rm i,z} \frac{1}{r} \partial_r (r \Pi_{r\zeta,j}).
\end{split}
\end{equation}
Here, $\delta E_r=-\partial_r\delta\phi$. $\rho_{\scriptscriptstyle \rm M}=\sum_{j={\rm i,z}}n_jm_j$ is the total ion mass density. The NP factor ${\epsilon_r}$ remains unchanged upon the addition of impurities for the mesoscale $k_r\Delta r_j\ll1$ considered here, which is consistent with the results found by Guo \cite{Guo2017NF}. Equation \eqref{eq:Er} recovers equation (96) in \cite{Zhang2020NF} when impurities are neglected. As in pure plasmas, the driving effects of the main-ion/impurity TEF on ZFs are not shielded by the NP effect in toroidal geometry, while the driving effects of the main-ion/impurity PRS are still shielded. Note that equation \eqref{eq:Er} is similar to equation (15) in \cite{Li2026POP}, in which the TEF and PRS are represented by the temperature and density perturbations, respectively.

Equation \eqref{eq:Er} shows that both main-ion and impurity turbulent transport fluxes can drive ZFs. More specifically, the first term on the right-hand side shows that the TEF drive contains the factor $m_{j}/e_{j}$. In the case of the fully stripped impurities with $m_{\rm z}/e_{\rm z}$ close to $m_{\rm i}/e_{\rm i}$, the difference between the impurity and main-ion TEF drive is mainly determined by their magnitudes. Therefore, the impurity TEF drive can be comparable to that of the main-ion TEF when the magnitude of $Q_{r,\rm z}$ is comparable to that of $Q_{r,\rm i}$. On the other hand, equations \eqref{eq:PRS} and \eqref{eq:TRS} show that the turbulent PRS and TRS are proportional to the mass density $n_{j}m_{j}$. Therefore, the impurity PRS (TRS) drive can also be comparable to that of the main-ion PRS (TRS) when $n_{\rm z}m_{\rm z}$ is comparable to $n_{\rm i}m_{\rm i}$.

Equation \eqref{eq:Er} describes temporal asymptotic ZFs and is valid on the timescale longer than several ion bounce periods $\tau_{{\rm b}}$. However, on the timescale shorter than $\tau_{{\rm b}}$, the turbulent PRS drive is not fully shielded \cite{Wang2025NF}. Following \cite{Wang2025NF}, we introduce the response function $g(\tau)$ to capture this time-dependent PRS drive, and the ZF driving equation is generalized to
\begin{equation}\label{eq:Er gt}
\begin{split}
\delta E_r(t) 
&= -\frac{1}{\rho_{\scriptscriptstyle \rm M}} \sum_{j=\rm i,z} \frac{m_j}{e_j} \int_{-\infty}^t {\rm d}t' \, \partial_r \left[ \frac{1}{r} \partial_r \left( r \frac{2}{3} Q_{r,j}(t') \right) \right] \\
&\quad + \frac{B_T}{\rho_{\scriptscriptstyle \rm M}} \sum_{j=\rm i,z} \int_{-\infty}^t {\rm d}t' \, \frac{1}{r} \partial_r \left[ r \Pi_{r \theta,j}(t') \right] \cdot g(t-t')\\
&\quad - \frac{B_P}{\rho_{\scriptscriptstyle \rm M}} \sum_{j=\rm i,z} \int_{-\infty}^t {\rm d}t' \, \frac{1}{r} \partial_r \left[ r \Pi_{r \zeta,j}(t') \right], 
\end{split}
\end{equation}
in which $g$ satisfies
\begin{equation}\label{eq:gt_limit}
g(\tau) =
\begin{cases} 
1, & \text{if } \tau/\tau_{{\rm b}} \to 0; \\
1 / \epsilon_r, & \text{if } \tau/\tau_{{\rm b}} \to \infty.
\end{cases}
\end{equation}
We point out that the response function $g(\tau)$ is independent of particle species, which will be demonstrated later.

\section{Simulation results}\label{sec:simulation}
In section \ref{sec:theory}, the driving equation of ZFs in multi-ion-species tokamak plasmas is derived by using the gyrokinetic theory. The driving effects of the impurity turbulent transport fluxes, as well as the time-dependent shielding of the turbulent PRS drive, are discussed. To verify these results, a set of gyrokinetic simulations are performed by using the NLT code \cite{Ye2016POP,Xu2017POP}.

\subsection{The global gyrokinetic code NLT} \label{subsec:NLT introduction}
NLT is a nonlinear global gyrokinetic code, which uses the numerical Lie-transform method \cite{Wang2012POP,Wang2013PRE,Wang2013POP,Wang2014POP} to solve the gyrokinetic Vlasov equation. The early electrostatic adiabatic-electron version \cite{Ye2016POP} was verified by the linear GAM and ITG test in the standard Cyclone Base Case (CBC). Then, the NLT code adopted the Fourier filter and the field-aligned coordinates $(r,\alpha,\theta)$ with $\alpha=q\theta-\zeta$ and $\zeta$ the toroidal angle, and was verified by the nonlinear ITG turbulence test \cite{Xu2017POP}. Recently, the NLT code has been extended to include a multi-ion-species module and applied to simulations of the ITG instability \cite{Tang2023AIPAdv} and the turbulent transport \cite{Tang2025PPCF} in multi-ion-species tokamak plasmas.

The gyrokinetic Vlasov equation used in the NLT code is
\begin{equation}\label{eq:GKE-NLT}
    \frac{\mathrm{d}F_j}{\mathrm{d}t}=\frac{\partial F_j}{\partial t}+\dot{\boldsymbol{X}}\cdot\nabla F_j+\dot{v}_{\parallel}\frac{\partial F_j}{\partial v_{\parallel}}=0.
\end{equation}
Here, $F_j(\boldsymbol{X},v_{\parallel},\mu,t)=F_{0,j}+\delta F_j$ is the gyrocenter distribution function for $j$-species, with $F_{0,j}$ and 
$\delta F_j$ being its equilibrium and perturbed parts, respectively.
$\dot{\boldsymbol{X}}=\{\boldsymbol{X},H_{0,j}+H_{1,j}\}$ and  $\dot{v}_{\parallel}=\{v_{\parallel},H_{0,j}+H_{1,j}\}$,  with $\{\cdot,\cdot\}$ denoting the Poisson bracket determined by the equilibrium fields. $H_{0,j}=m_jv_{\parallel}^2/2+\mu B$ and $H_{1,j}=e_j\left\langle\delta\Phi\right\rangle_{{\rm GA}}$ are the gyrocenter equilibrium and perturbed Hamiltonian, respectively. $\delta\Phi$ is the perturbed electrostatic potential. $\langle \cdot\rangle_{{\rm GA}}$ denotes the gyroaverage.

The adiabatic electron model is adopted in the simulations, and the non-axisymmetric part of perturbed electrostatic potential $\delta\Phi$ is determined from the gyrokinetic quasi-neutrality equation \cite{Xu2026RMPP},

\begin{equation}\label{eq:GKQNE-NLT}
\sum_{j=\rm i,z} e_j(\delta n_{j, \text{pol}} + \delta n_{j, \text{gy}}) 
-n_{0,\rm e} \frac{e_{\rm e}^2\delta \Phi }{T_{\rm e}} = 0.
\end{equation}
Here, $\delta n_{j, \text{gy}}$ is the ion gyrocenter density, which is expressed as
\begin{equation}\label{eq:deltani_gy-NLT}
\delta n_{j, \text{gy}} = \int \mathrm{d}^3 \bm v \, \langle \delta F_j \rangle_{\text{GA}},
\end{equation}
$\delta n_{j, \text{pol}}$ is the ion polarization density, which is expressed as
\begin{equation}
\delta n_{j,\rm pol} = -\frac{n_{0,j}e_j}{T_j} \left( \delta \Phi - \int {\rm d}^3\bm v \frac{F_{0,j}}{n_{0,j}} \langle \langle \delta \Phi \rangle_{\rm GA} \rangle_{\rm GA} \right).
\end{equation}
The zonal part of perturbed electrostatic potential $\delta\Phi$ is solved by the gyrokinetic quasi-neutrality equation in the long-wavelength limit,
\begin{equation}\label{eq:QNE-NLT}
    \nabla\cdot\left(\sum_{j={\rm i,z}}\frac{n_{j}m_j}{B^2}\nabla_{\perp}\delta\Phi\right)-n_{0,\rm e} \frac{e_{\rm e}^2(\delta \Phi - \langle \delta \Phi \rangle_{\rm FA})}{T_{\rm e}}
    =-\sum_{j={\rm i,z}}e_j\delta n_{j,{\rm gy}}.
\end{equation}

\subsection{Linear benchmark for the ITG instability in the presence of impurities}\label{subsec:verification}
\begin{figure}[htbp]
\centering
\includegraphics[width=0.6\textwidth]{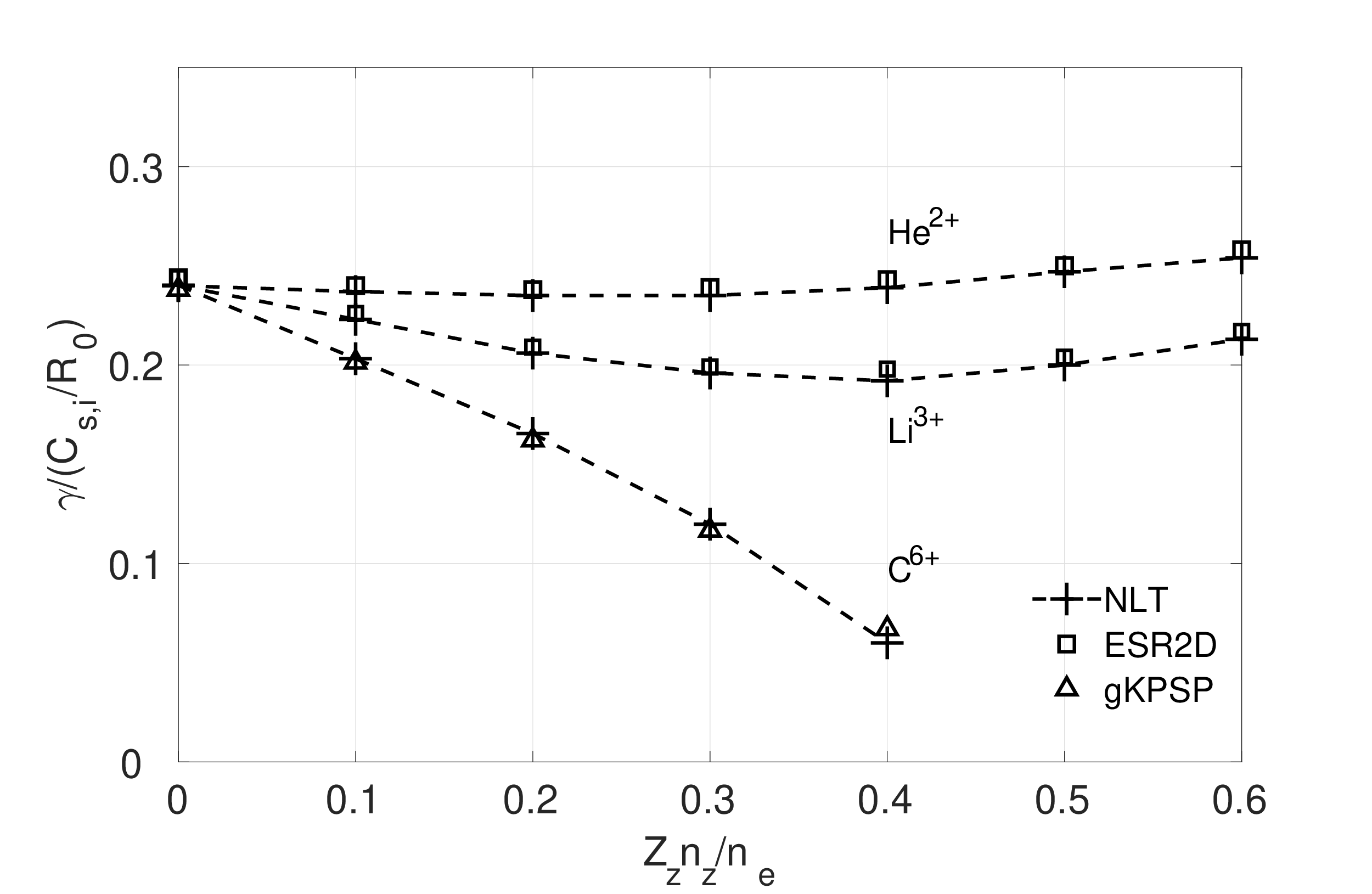}
\caption{The linear growth rates of the ITG instability for He$^{2+}$, Li$^{3+}$, and C$^{6+}$ at different impurity charge concentration $Z_{\rm z}n_{\rm z}/n_{\rm e}$. The linear growth rate $\gamma$ is normalized by $C_{\rm s,i}/R_0$.}
\label{fig:linearITG}
\end{figure}

The linear growth rates of the ITG instability for different impurity species and charge concentrations $Z_{\rm z}n_{\rm z}/n_{\rm e}$ are calculated by using the NLT code. The NLT results are benchmarked against other global gyrokinetic codes, e.g., the global initial-value codes gKPSP \cite{Seo2022POP} and the global eigenvalue code ESR2D \cite{Wanghc2026NF}, as shown in figure \ref{fig:linearITG}. In this linear benchmark, deuterium (D$^{+}$) is the main ion, and the fully stripped helium (He$^{2+}$), lithium (Li$^{3+}$), and carbon (C$^{6+}$) are the impurities. At $r=0.5a$, $q=1.4$, $R/L_{T_{\rm i}}=R/L_{T_{\rm z}}=R/L_{T_{\rm e}}=6.9$, $R/L_{n_{\rm i}}=R/L_{n_{\rm z}}=2.2$, $k_\theta\rho_{\rm i}=0.3$. $k_\theta$ is the poloidal wave number, $\rho_{\rm i}=\sqrt{m_{\rm i}T_{\rm i}}/e_{\rm i}B$, and $L_A=-A/({\rm d}A/{\rm d}r)$ is the scale length. The NLT results agree well with those of other gyrokinetic codes. For C$^{6+}$, the growth rate $\gamma$ varies rapidly with $Z_{\rm z}n_{\rm z}/n_{\rm e}$, whereas for He$^{2+}$ and Li$^{3+}$, it remains nearly unchanged. Qualitatively similar results were also found by Dong \cite{Dong1994AIP}.

\subsection{The nonlinear simulation setup}\label{subsec:Simulation setup}

\begin{figure}[htbp]
\centering
\includegraphics[width=0.6\textwidth]{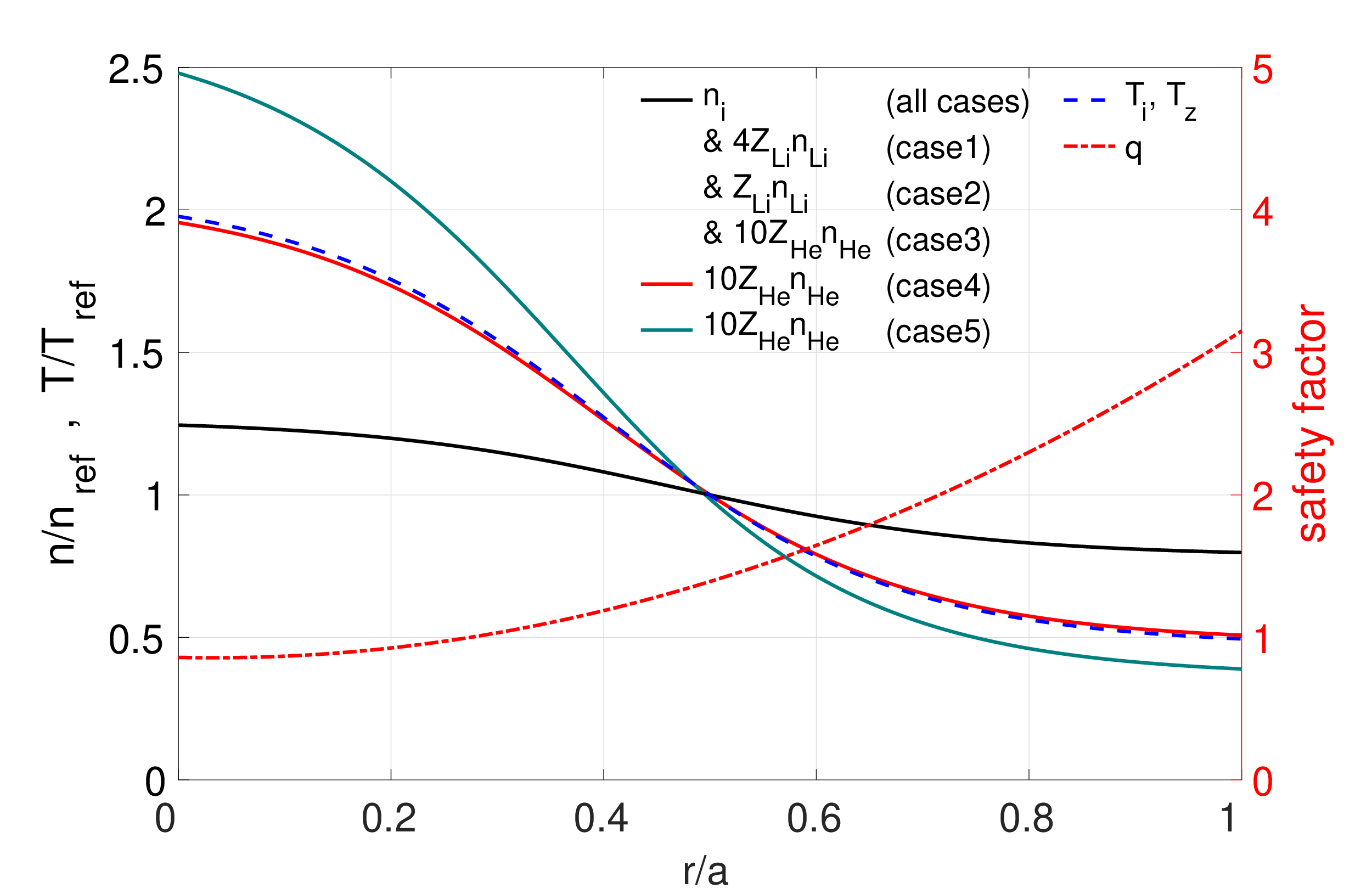}
\caption{Density, temperature and safety factor profiles in the CBC parameters. Black solid line denotes the main ion density profiles in all cases, or the scaled impurity charge density profiles in cases 1-3. The red and teal solid lines denote the scaled impurity charge density profiles in cases 4 and 5, respectively. $n_{\rm ref}=10^{19}~\text{m}^{-3}$ and $T_{\rm ref}=1.97~\text{keV}$.}
\label{fig:profile}
\end{figure}

\begin{table}[htbp]
\centering
\setlength{\tabcolsep}{15pt}  
\caption{The impurity charge concentration $Z_{\rm z}n_{\rm z}/n_{\rm e}$, effective charge number $Z_{\rm eff}$, and the ratio of density gradient $L_{n_{\rm i}}/L_{n_{\rm z}}$ at $r=0.5a$ for the different nonlinear simulation cases. $n_{\rm i}=10^{19}\ \rm m^{-3}$ and $R/L_{n_{\rm i}}=2.2$ at $r=0.5a$. $n_{\rm i}$ and $L_{n_{\rm i}}(r=0.5a)$ are kept the same in all cases, thus the higher $L_{n_{\rm i}}/L_{n_{\rm z}}$ indicates the steeper impurity density gradient. The $n_{\rm e}$ profile changes with $L_{n_{\rm i}}/L_{n_{\rm z}}$ due to the quasi-neutrality condition. Note that the growth rates of the linear ITG mode in cases 1 and 2 are almost the same, which can be seen in figure \ref{fig:linearITG}.}
\begin{tabular}{ccccc}
\hline\hline  
 & Impurity & $Z_{\rm z}n_{\rm z}/n_{\rm e}$ & $Z_{\rm{eff}}$ & $L_{n_{\rm i}}/L_{n_{\rm z}}$ \\
\midrule
case 1 & Li$^{3+}$ & $\bm{0.2}$ & $\bm{1.4}$ & 1 \\
case 2 & Li$^{3+}$ & $\bm{0.5}$ & $\bm{2}$ & 1 \\
case 3 & He$^{2+}$ & 0.1 & 1.1 & $\bm{1}$ \\
case 4 & He$^{2+}$ & 0.1 & 1.1 & $\bm{3}$ \\
case 5 & He$^{2+}$ & 0.1 & 1.1 & $\bm{5}$ \\
\bottomrule
\end{tabular}
\label{tab:impurity parameter}
\end{table}

A set of nonlinear ITG relaxation simulations in the presence of impurities are performed to examine the driving mechanisms of ZFs in multi-ion-species tokamak plasmas. The CBC parameters are used, with the major radius $R_0=1.67~\rm m$, the minor radius $a=0.6~\rm m$, and the magnetic field on axis $B_0=1.9~\rm T$. Ion equilibrium density $n_j$, temperature $T_j$ and safety factor $q$ profiles are shown in figure \ref{fig:profile}. As mentioned in section \ref{sec:introduction}, the impurity charge concentration $Z_{\rm z}n_{\rm z}/n_{\rm e}$ and impurity density gradient $R/L_{n_{\rm z}}$ affect the turbulent transport flux \cite{Romanelli2011NF,Romanelli2013NF}, which may thus affect ZFs. To disentangle these two effects, five cases with different impurity charge concentration and impurity density gradient are designed, as shown in table \ref{tab:impurity parameter}. Cases 1 and 2 vary the Li$^{3+}$ charge concentration $Z_{\rm z}n_{\rm z}/n_{\rm e}$ at a fixed impurity density gradient, whereas cases 3-5 vary the He$^{2+}$ density gradient $R/L_{n_{\rm z}}$ at a fixed impurity charge concentration at $r=0.5a$. Note that the high Li$^{3+}$ charge concentration in case 2 is motivated by the initial phase of an FTU discharge with the liquid lithium limiter \cite{Mazzitelli2011NF}, in which $Z_{\rm Li}n_{\rm Li}/n_{\rm e}$ is up to 50\%. The high He$^{2+}$ density gradient in case 5 is motivated by the helium ash profile in the ITER simulation \cite{Budny2008NF}.  $n_{\rm i}$ and $T_{\rm i}$ are kept the same in all cases. $T_{\rm i}=T_{\rm z} = T_{\rm e}$ is assumed.

The simulation domain is $r/a\in[0,1]$, $\alpha\in[0,\pi]$, $\theta\in[-\pi,\pi]$, $v_{\parallel}/C_{{\rm s},j}\in[-4,4]$, $\mu B_0/T_{j}\in[0,4^2/\sqrt{2}]$. The grid numbers are $N_r\times N_\alpha\times N_\theta\times N_{v_\parallel}\times N_\mu=222\times 190\times 16\times 64\times16$ with Gauss-Laguerre discretization \cite{Xiao2025PST} on the $\mu$ grid and uniform discretization on others. The simulation time step is $\Delta t=2\tau_{{\rm i},{\rm gy}}$, with $\tau_{{\rm i},{\rm gy}}$ being the main-ion gyroperiod at the magnetic axis. A damping buffer region is adopted near the radial outer boundary. A mean value theorem \cite{Wu2026PST} is used to obtain the perturbed distribution function and electrostatic potential at the magnetic axis.

\subsection{The turbulent energy flux in the presence of impurities}\label{subsubsec:TEF}

\begin{figure}[htbp]
\centering
\includegraphics[width=0.7\textwidth]{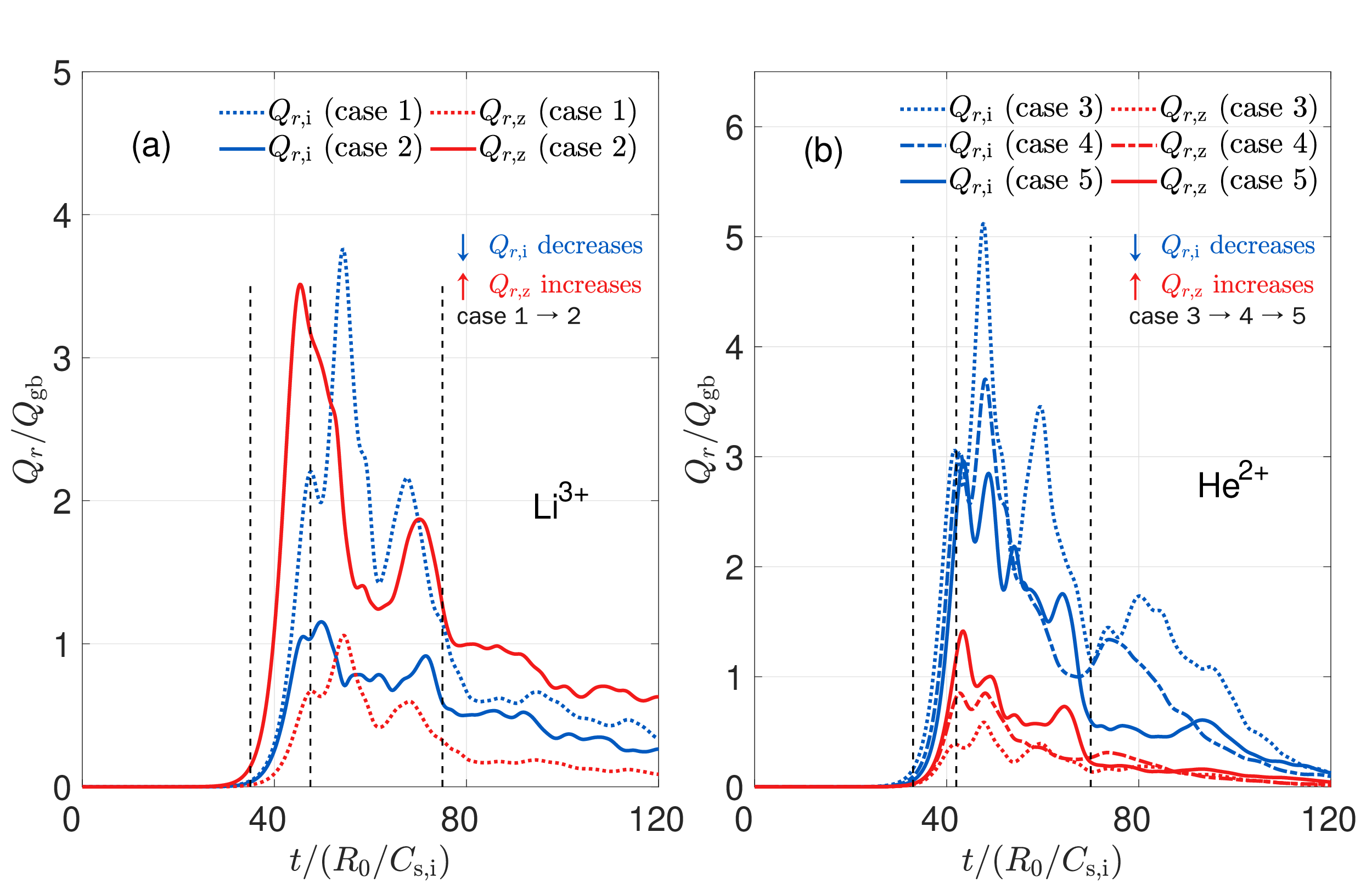}
\caption{Main-ion TEF $Q_{r,\rm i}$ (blue lines) and impurity TEF $Q_{r,\rm z}$ (red lines) in (a) cases 1, 2 and (b) cases 3, 4, 5. The TEF shown here is radially averaged over $r=0.4\sim0.6a$. $Q_{\rm gb}=n_{\rm ref}T_{\rm ref}C_{\rm s,i}\rho_{\rm  i}^2/a^2$. The three black dashed vertical lines from left to right mark the quasilinear, saturation, and nonlinear stages, respectively.}
\label{fig:Qr}
\end{figure}

Figure \ref{fig:Qr} shows the main-ion and impurity TEF in cases 1 and 2 [figure \ref{fig:Qr}(a)] and cases 3-5 [figure \ref{fig:Qr}(b)]. The impurity TEF $Q_{r,\rm z}$ increases, whereas the main-ion TEF $Q_{r,\rm i}$ decreases as $Z_{\rm z}n_{\rm z}/n_{\rm e}$ or $L_{n_{\rm i}}/L_{n_{\rm z}}$ increases. In case 2 with the high-charge-concentration Li$^{3+}$ impurity, $Q_{r,\rm z}$ is even larger than $Q_{r,\rm i}$ [figure \ref{fig:Qr}(a)], whereas in case 5 with the low-charge-concentration but steep-density-gradient helium ash, $Q_{r,\rm z}$ is still comparable to $Q_{r,\rm i}$ [figure \ref{fig:Qr}(b)].

\indent To understand the opposite behavior between $Q_{r,\rm z}$ and $Q_{r,\rm i}$ as $Z_{\rm z}n_{\rm z}/n_{\rm e}$ and $L_{n_{\rm i}}/L_{n_{\rm z}}$ vary, the TEF $Q_{r,j}$ is decomposed into the turbulent heat flux $q_{r,j}$ and the turbulent particle flux $\Gamma_{r,j}$, 
\begin{equation}\label{eq:Qr}
    Q_{r,j} = q_{r,j}+\frac{3}{2}T_j\Gamma_{r,j}.
\end{equation}

\begin{figure}[htbp]
\centering
\includegraphics[width=0.9\textwidth]{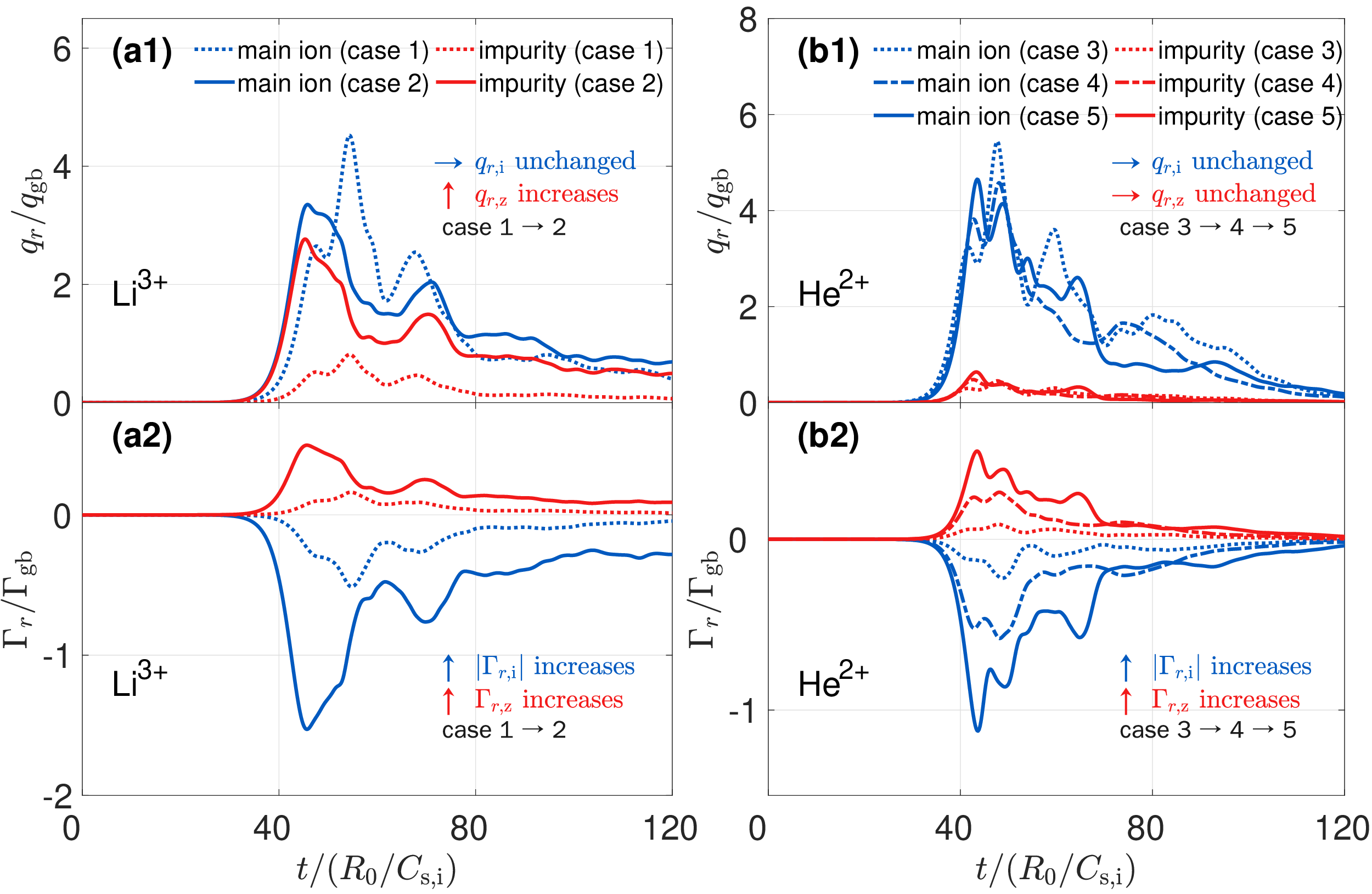}
\caption{Turbulent heat flux $q_{r,j}$ [(a1),(b1)] and turbulent particle flux $\Gamma_{r,j}$ [(a2),(b2)] of main ions (blue lines) and impurities (red lines) in cases 1-5. Turbulent heat and particle fluxes shown here are averaged on $r=0.4\sim0.6a$. $q_{\rm gb}=Q_{\rm gb}$, $\Gamma_{\rm gb}=n_{\rm ref}C_{\rm s,i}\rho_{\rm  i}^2/a^2$.}
\label{fig:gammar_qr}
\end{figure}

\indent The resulting $q_{r,j}$ and $\Gamma_{r,j}$ are shown in figure \ref{fig:gammar_qr}. For impurities, as $Z_{\rm z}n_{\rm z}/n_{\rm e}$ increases, both the outward $q_{r,\rm z}$ and $\Gamma_{r,\rm z}$ increase [figure \ref{fig:gammar_qr}(a1), (a2)], so that $Q_{r,\rm z}$ increases with $Z_{\rm z}n_{\rm z}/n_{\rm e}$. While as $L_{n_{\rm i}}/L_{n_{\rm z}}$ increases, the outward $q_{r,\rm z}$ remains nearly unchanged [figure \ref{fig:gammar_qr}(b1)], but the outward $\Gamma_{r,\rm z}$ increases [figure \ref{fig:gammar_qr}(b2)], so that $Q_{r,\rm z}$ also increases with $L_{n_{\rm i}}/L_{n_{\rm z}}$. For main ions, $q_{r,\rm i}$ remains nearly unchanged as either $Z_{\rm z}n_{\rm z}/n_{\rm e}$ or $L_{n_{\rm i}}/L_{n_{\rm z}}$ increases [figure \ref{fig:gammar_qr}(a1), (b1)]. However, the quasi-neutrality condition, $Z_{\rm i}\Gamma_{r,\rm i}=-Z_{\rm z}\Gamma_{r,\rm z}$, requires an inward $\Gamma_{r,\rm i}$ whose magnitude increases with $Z_{\rm z}n_{\rm z}/n_{\rm e}$ and $L_{n_{\rm i}}/L_{n_{\rm z}}$ [figure \ref{fig:gammar_qr}(a2), (b2)]; this inward $\Gamma_{r,\rm i}$ increasingly counteracts the contribution of outward $q_{r,\rm i}$ to $Q_{r,\rm i}$, so that $Q_{r,\rm i}$ decreases with $Z_{\rm z}n_{\rm z}/n_{\rm e}$ and $L_{n_{\rm i}}/L_{n_{\rm z}}$.

The opposite behaviors of \(Q_{r,\rm i}\) and \(Q_{r,\rm z}\) thus mainly originate from the quasi-neutrality condition, which makes \(\Gamma_{r,\rm z}\) and \(\Gamma_{r,\rm i}\) increase in opposite directions. Furthermore, the increase of $\Gamma_{r,\rm z}$ with $Z_{\rm z}n_{\rm z}/n_{\rm e}$ and $L_{n_{\rm i}}/L_{n_{\rm z}}$ can be understood from the impurity turbulent transport theory by Angioni \cite{Angioni2021PPCF}, in which $\Gamma_{r,\rm z}$ is decomposed as

\begin{equation}\label{eq:angioni}
\frac{R \Gamma_{r,\rm z}}{n_{\rm z}} = D_{n,\rm z} \frac{R}{L_{n_{\rm z}}} + D_{T,\rm z} \frac{R}{L_{T_{\rm z}}}+ D_{u,\rm z}u_{\rm z}' + RV_{p,\rm z}.
\end{equation}
Here, $D_{n,\rm z}$, $D_{T,\rm z}$, $D_{u,\rm z}$, and $V_{p,\rm z}$ represent impurity turbulent diffusion, thermo-diffusion, roto-diffusion, and pure convection, respectively. $u_{\rm z}'$ is the normalized toroidal rotation gradient. The roto-diffusion term $D_{u,\rm z}u_{\rm z}'$ can be ignored because there is no equilibrium toroidal rotation in our simulations. In the strong-ballooning limit, and for $R/L_{n_{\rm z}}\approx2$ as realized in cases 1 and 2, the turbulent diffusion $D_{n,\rm z} R/L_{n_{\rm z}}$ is almost canceled by the pure convection $RV_{p,\rm z}$ \cite{Angioni2021PPCF,Angioni2009NF,Wang2016POP}. This cancellation is also found in the quasilinear calculation of the deuterium–tritium particle fluxes \cite{Tang2023AIPAdv}. Thus $\Gamma_{r,\rm z}/n_{\rm z}$ is almost determined by the thermo-diffusion term $D_{T,\rm z} R/L_{T_{\rm z}}$, which is assumed to be the same in cases 1 and 2. Therefore, $\Gamma_{r,\rm z}$ is proportional to $n_{\rm z}$ and increases with $Z_{\rm z}n_{\rm z}/n_{\rm e}$, consistent with the simulation results [figure \ref{fig:gammar_qr}(a2)]. For cases 3-5, an increase in $L_{n_{\rm i}}/L_{n_{\rm z}}$ leads to a larger turbulent diffusion term $D_{n,\rm z}R/L_{n_{\rm z}}$, and hence a larger $\Gamma_{r,\rm z}/n_{\rm z}$. Since $n_{\rm z}$ in the radial region $0.4a\sim0.6a$ are almost the same for cases 3-5, $\Gamma_{r,\rm z}$ increases with $L_{n_{\rm i}}/L_{n_{\rm z}}$, which is also consistent with the simulation results [figure \ref{fig:gammar_qr}(b2)].

Here, we give an explanation for the direction of the main-ion and impurity turbulent particle fluxes (i.e., $\Gamma_{r,\rm i}<0$ and $\Gamma_{r,\rm z}>0$) observed in all cases 1–5. The direction of the thermo-diffusion term in equation \eqref{eq:angioni} depends on the sign of ($E_{{\rm res},j}/T_j-3/2$), with $E_{{\rm res},j}$ denoting the resonant energy of the magnetic drift resonance \cite{Tang2023AIPAdv,Angioni2021PPCF}. $E_{{\rm res},j}$ is proportional to $Z_j$ \cite{Angioni2021PPCF}, so that $E_{\rm res,i}/T_{\rm i}<3/2<E_{\rm res,z}/T_{\rm z}$ at $T_{\rm i}=T_{\rm z}$, which indicates an inward main-ion and an outward impurity thermo-diffusion. A similar result was found in \cite{Tang2023AIPAdv}, where the outward tritium and inward deuterium thermo-diffusion arise from the same $E_{{\rm res},j}$ but different temperatures ($T_{\rm D} > T_{\rm T}$), giving $E_{\rm res,D}/T_{\rm D}<3/2<E_{\rm res,T}/T_{\rm T}$. In the present work, the outward impurity thermo-diffusion dominates the outward $\Gamma_{r,\rm z}$ in cases 1-3 ($R/L_{n_{\rm z}}\approx2$). For cases 4-5 with the steep-density-gradient helium ash ($R/L_{n_{\rm z}}>2$), the outward impurity turbulent diffusion $D_{n,\rm z}R/L_{n_{\rm z}}$ is larger than the inward convection $RV_{p,\rm z}$, also leading to an outward $\Gamma_{r,\rm z}$. Note that the helium ash produced by fusion reactions generally satisfies $R/L_{n_{\rm z}}>2$ and $T_{\rm z}\approx T_{\rm i}$ \cite{Synakowski1995PRL,Budny2008NF}, thus the theoretical analysis and simulation results shown here indicate that ITG turbulence could provide a radially outward transport channel for helium ash in a steep density gradient.

Note that similar behavior of $\Gamma_{r,j}$ in our work has also been observed in local gyrokinetic simulations by the GKW code \cite{Romanelli2011NF,Romanelli2013NF} and the GENE code \cite{Angioni2009NF}.
For deuterium plasmas with a core-peaked and high-charge-concentration lithium profile, an inward turbulent particle flux for deuterium and an outward flux for lithium, both driven by ITG turbulence, are reported in \cite{Romanelli2011NF,Romanelli2013NF}. Furthermore, the magnitudes of these two particle fluxes both increase with the lithium charge concentration and its density gradient. In \cite{Angioni2009NF}, the thermo-diffusion of helium turbulent particle flux was found to be outward, and the magnitude of helium ash turbulent particle flux significantly increases with the helium ash density gradient. These local simulation results are qualitatively consistent with those in our global simulations.

\subsection{The role of impurities in the ZF driving mechanisms}\label{subsubsec:ZREF}
\begin{figure}[htbp]
\centering
\includegraphics[width=0.7\textwidth]{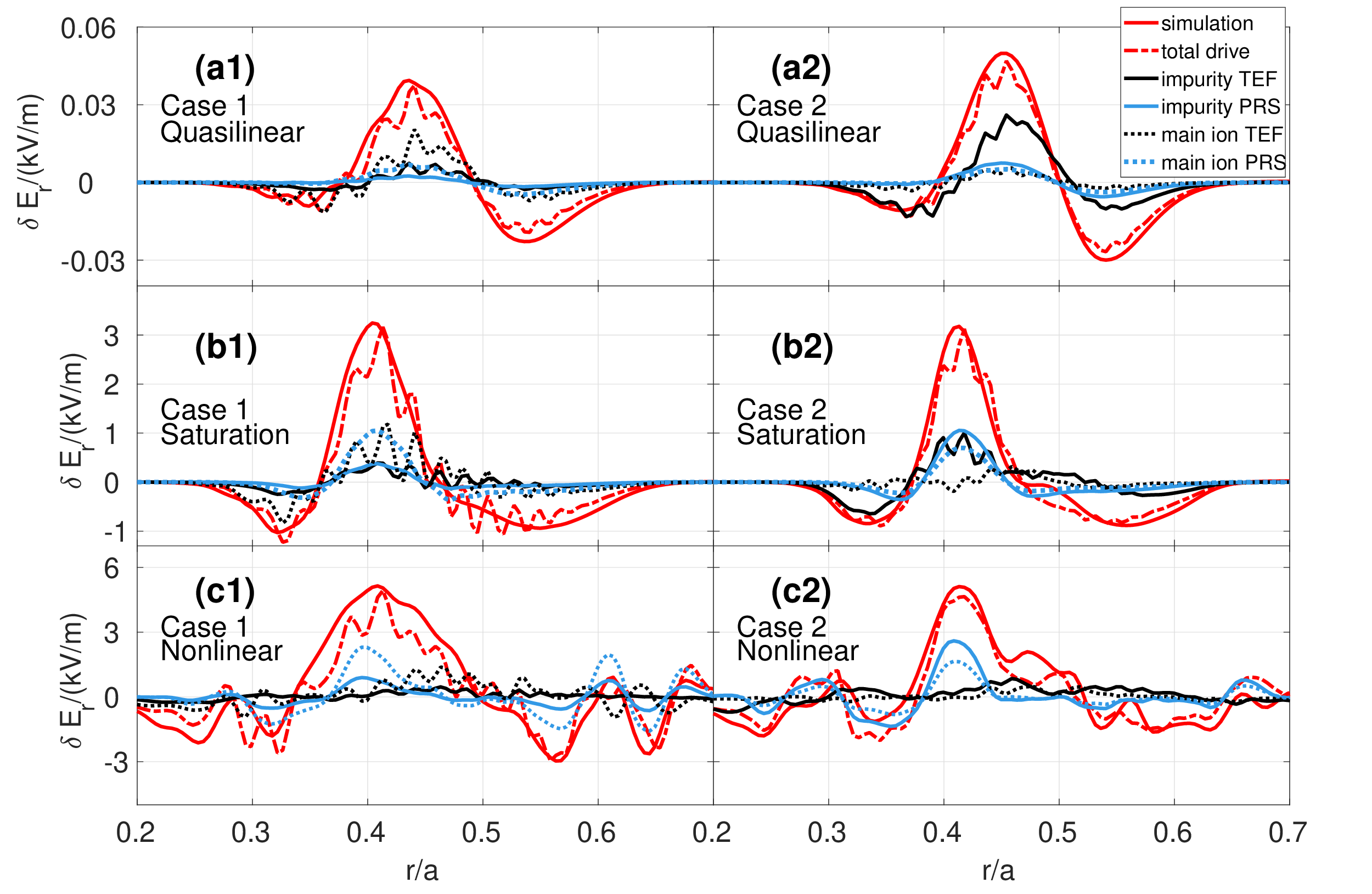}
\caption{Simulation results of ZFs (red solid lines) in the quasilinear, saturation and nonlinear stages for cases 1 and 2. The selection of these three stages is shown in figure \ref{fig:Qr}. The theoretical predictions from equation \eqref{eq:Er gt} are shown in red dash-dotted lines. The driving effects of impurity TEF and PRS in equation \eqref{eq:Er gt} are shown in black and blue solid lines, respectively; while those of main ion TEF and PRS are shown in black and blue dotted lines, respectively. The TEF contributions in the nonlinear stage are calculated following the method in \cite{Wang2025NF}.}
\label{fig:Er_Zznz}
\end{figure}

\begin{figure}[htbp]
\centering
\includegraphics[width=0.7\textwidth]{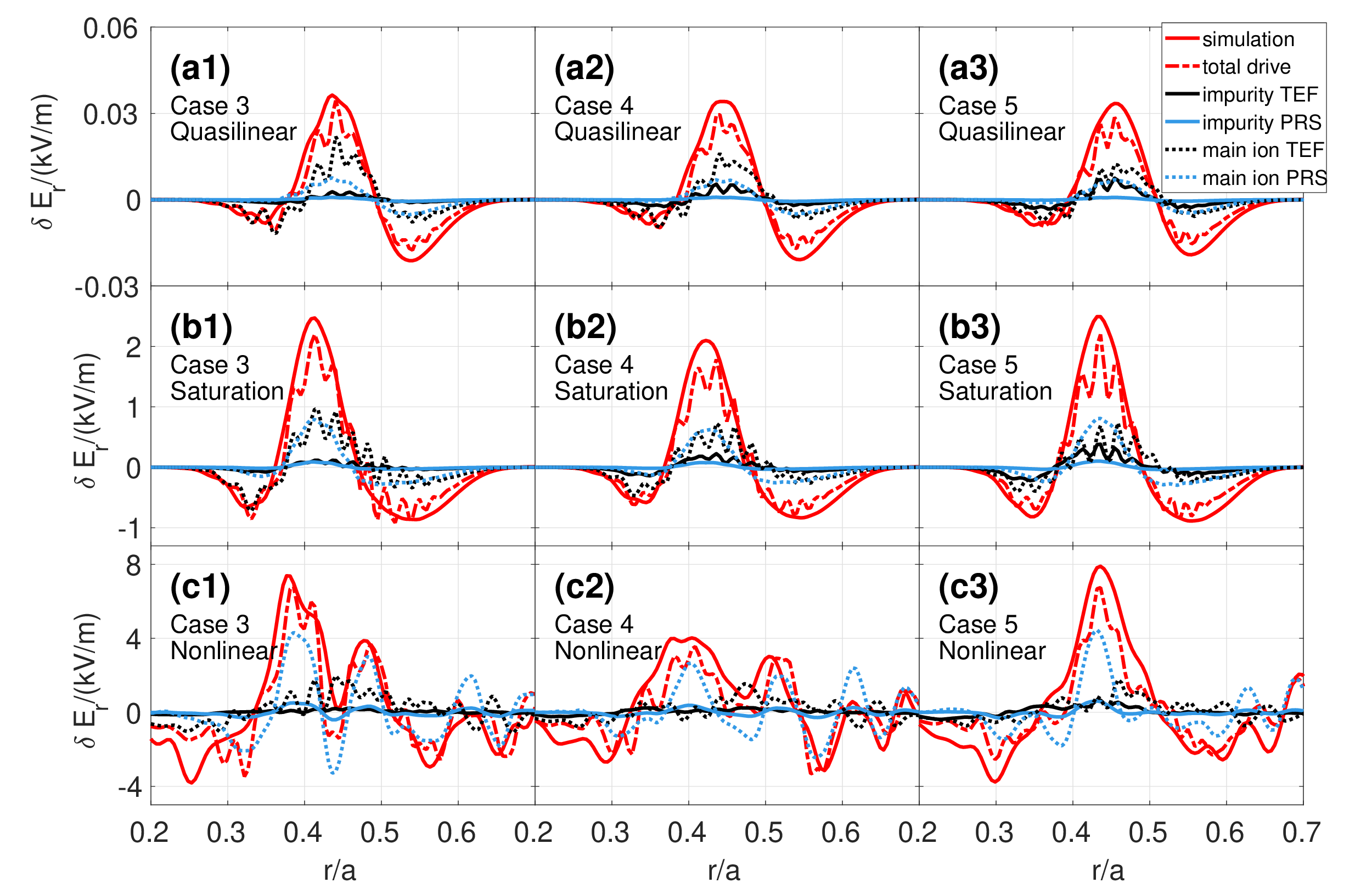}
\caption{Same as figure \ref{fig:Er_Zznz}, but for cases 3-5.}
\label{fig:Er_R_Lnz}
\end{figure}

To verify the gyrokinetic theory in section \ref{sec:theory}, the NLT simulation results of ZFs and the theoretical predictions from equation \eqref{eq:Er gt} are compared in figures \ref{fig:Er_Zznz} and \ref{fig:Er_R_Lnz}. The theoretical predictions (red dash-dotted lines) agree well with the simulation results (red solid lines). Table \ref{tab:TEFandPRS} lists the contributions of the main-ion and impurity TEF and PRS to ZFs. Note that the contributions of TRS are relatively small for all cases, and are therefore not shown but considered in the total drive of ZFs (red dash-dotted lines). $k_r\Delta r_i\approx 0.5$ is estimated from these simulation results, so that the mesoscale assumption is valid.

\begin{table}[htbp]
\centering
\caption{The contributions of the main-ion and impurity TEF and PRS to the simulated ZF by equation \eqref{eq:Er gt}. The sum of these four drives are listed. The contribution of the turbulent TRS is small, and not shown in this table. Bold numbers mean the dominant drive.}
\begin{tabular}{ccccccc}
\toprule
 &  & Impurity TEF & Main ion TEF & Impurity PRS & Main ion PRS &   Sum   \\
\midrule
\multicolumn{1}{c|}{\multirow{3}{*}{case 2}} & Quasilinear  & $\bm{52\%}$ & 11\%  & 15\% & 10\% & 88\% \\
\multicolumn{1}{c|}{}                        & Saturation   & $31\%$ & 3\% & $\bm{33\%}$ & 22\% & 89\% \\
\multicolumn{1}{c|}{}                        & Nonlinear    & 5\%   & 1\%  & $\bm{50\%}$ & 32\% & 88\% \\
\midrule
\multicolumn{1}{c|}{\multirow{3}{*}{case 5}} & Quasilinear  & 22\%  & $\bm{37\%}$ & 2\%  & 21\% & 82\% \\
\multicolumn{1}{c|}{}                        & Saturation   & 15\%  & 28\% & 4\%  & $\bm{33\%}$ & 80\% \\
\multicolumn{1}{c|}{}                        & Nonlinear    & 9\%   & 11\%  & 8\% & $\bm{56\%}$ & 84\% \\
\bottomrule
\end{tabular}
\label{tab:TEFandPRS}
\end{table}

We first focus on the TEF drive (black lines). As discussed in section \ref{sec:theory}, the TEF drive is weighted by the mass-to-charge ratio $m_j/e_j$. Because $m_{\rm z}/e_{\rm z}$ for Li$^{3+}$ and He$^{2+}$ is close to $m_{\rm i}/e_{\rm i}$ for D$^{+}$, the relative TEF drives of main ions and impurities are determined by the relative magnitude of $Q_{r,j}$. In section \ref{subsubsec:TEF}, we show that the magnitude of the impurity TEF $Q_{{r,\rm z}}$ is even larger than that of the main-ion TEF $Q_{{r,\rm i}}$ in case 2, while in case 5, it is comparable to that of $Q_{{r,\rm i}}$. Therefore, we can see from figures \ref{fig:Er_Zznz}(a2)-(b2) that $Q_{{r,\rm z}}$ (black solid lines) drives ZFs more strongly than $Q_{{r,\rm i}}$ (black dotted lines) in case 2. In case 5, $Q_{{r,\rm z}}$ and $Q_{{\rm r,i}}$ drive ZFs comparably [figures \ref{fig:Er_R_Lnz}(a3)-(b3)]. The contribution of impurity TEF to ZFs in case 5 can reach 22\% and 15\% in the quasilinear and saturation stages, respectively, which indicates that the impurity TEF drive cannot be ignored in the case with high-density-gradient helium ash, even though the helium ash charge concentration is low.

Then, we focus on the turbulent PRS drive. Figures \ref{fig:Er_Zznz}(a2)-(c2) shows that the impurity turbulent PRS drive (blue solid line) in case 2 is stronger than that of the main-ion turbulent PRS (blue dashed line). We recall that equations \eqref{eq:PRS} and \eqref{eq:Er gt} show that the ratio of impurity turbulent PRS drive to main-ion one is approximately equal to the mass-density ratio $n_{\rm z}m_{\rm z}/n_{\rm i}m_{\rm i}$. Table \ref{tab:impurity parameter} shows that $n_{\rm z}m_{\rm z}/n_{\rm i}m_{\rm i}$ in case 2 is approximately 1.2. Thus, the larger impurity turbulent PRS drive shown in figures \ref{fig:Er_Zznz}(a2)-(c2) is primarily a consequence of the larger impurity mass density. Note that the simulated ratios of the impurity to main-ion turbulent PRS drive shown in figures \ref{fig:Er_Zznz}-\ref{fig:Er_R_Lnz} and table \ref{tab:TEFandPRS} are slightly larger than the theoretical prediction $n_{\rm z}m_{\rm z}/n_{\rm i}m_{\rm i}$. This is because $\tilde {V}_{r}$ and $\tilde {V}_{\theta}$ in equation \eqref{eq:PRS} are gyro-averaged with $\rho_{{\rm \scriptscriptstyle L,i}}>\rho_{{\rm \scriptscriptstyle L,z}}$.

Table \ref{tab:TEFandPRS} shows that the turbulent PRS drive is predominant 
in the nonlinear stage. As noted in section \ref{sec:theory}, the turbulent PRS drive is time-dependently shielded by the NP effects, which is described by the response
function satisfying equation \eqref{eq:gt_limit}. However, the shielding timescale of the turbulent PRS drive has not been quantitatively determined. Here, we propose that this shielding timescale can be measured by the standard linear GAM test. The response function is then written as
\begin{equation}\label{eq:gt}
    g(\tau) = \frac{1}{\epsilon_r}+(1-\frac{1}{\epsilon_r}){\rm e}^{-\gamma_G\tau}.
\end{equation}
$\gamma_G$ is the damping rate of GAM in multi-ion-species tokamak plasmas \cite{Guo2010POP}. Note that the turbulent PRS is non-ambipolar, whose driving effects can be represented by the density perturbation in the GAM test. Using the parameters of cases 1-5, the GAM test results show that $\gamma_G/(C_{\rm s,i}/R_0)=0.26, 0.25, 0.23, 0.23$, and $0.24$ for cases 1-5, respectively, and agree with the GAM theory by Guo \cite{Guo2010POP}, which predicts that $\gamma_G$ is almost unchanged when $Z_{\rm eff}<4$ in the deuterium plasma. It is also found that $\gamma_G$ is almost the same for the GAM excited by either main-ion or impurity density perturbation, which indicates that the time-dependent PRS drives of both the main ions and impurities are described by the same response function $g(\tau)$. The turbulent PRS drives in figures \ref{fig:Er_Zznz} and \ref{fig:Er_R_Lnz} are all calculated with response function $g(\tau)$ given by equation \eqref{eq:gt}. In addition, for cases 1-5 and an additional high-carbon-charge-concentration case where $\gamma_G/(C_{\rm s,i}/R_0)=0.17$, the theoretical predictions obtained with equation \eqref{eq:gt}, $g(\tau)=1$, and $g(\tau)=1/\epsilon_r$ are compared in \ref{appen:gt}; the latter two response functions represent the unshielded and totally shielded turbulent PRS drive, respectively. It is found that the theoretical predictions obtained with equation \eqref{eq:gt} are consistent with the NLT simulation results in all cases, which indicates that the shielding timescale of the turbulent PRS drive is the damping time of GAM.

The turbulent PRS drive is shielded by the NP effect, which is mainly due to trapped ions \cite{RH1998PRL}, and thus the shielding timescale of the turbulent PRS drive is related to the ion bounce period $\tau_{{\rm b},j}$. The GAM test shown here provides a tool for quantitatively measuring this shielding timescale, with $1/\gamma_G\approx0.2\tau_{{\rm b,i}}$ for main ions ($\approx0.11\tau_{{\rm b,z}}$ for lithium and $\approx0.14\tau_{{\rm b,z}}$ for helium) in impurity scenarios considered in this work, which is significantly shorter than one ion bounce period $\tau_{{\rm b,i(z)}}$. Note that equation \eqref{eq:Er} describes the temporal asymptotic ZF with the ordering $\partial_t\ll 1/\tau_{{\rm b},j}$, rather than a time-dependent shielding process discussed here.

\section{Summary and discussion}\label{sec:summary}
In summary, by using gyrokinetic theory and simulations, we investigated the driving mechanisms of ZFs in ITG turbulence in the presence of impurities. The driving equation of ZFs in multi-ion-species tokamak plasmas is derived, which shows that both the impurity and main-ion turbulent transport fluxes drive ZFs significantly in certain scenarios. On this basis, a set of nonlinear simulations are performed for two physically distinct impurity scenarios, i.e., the core-peaked high-charge-concentration Li$^{3+}$ impurity and the low-concentration but steep-density-gradient helium ash. The simulation results for the ZF structure and amplitude agree well with the theoretical predictions. Two main findings are as follows. First, a large outward impurity turbulent particle flux and a large inward main-ion turbulent particle flux due to the quasi-neutrality condition are found in both impurity scenarios, which induce the magnitude of the impurity TEF comparable to or even stronger than that of the main-ion TEF, and thus the impurity TEF drives ZFs significantly in the quasilinear and saturation stages of ITG turbulence. Second, both the impurity and main-ion turbulent PRS drives are significant in the nonlinear stage for the case with a  high lithium charge concentration, and both of them are shielded by the time-dependent NP effects. The shielding timescale of the turbulent PRS drive is found to be the damping time
of GAM, which is significantly shorter than one ion bounce period.

In \cite{Kim2017POP} and \cite{Seo2022POP}, the global nonlinear gyrokinetic simulations show that impurities have effects on the ZF structure and amplitude. The change in ZF amplitude is attributed to the suppression of ITG turbulence in the presence of impurities \cite{Kim2017POP}. The ZF driving mechanisms investigated in this work may help clarify the mechanisms underlying impurity-induced changes in the ZF structure and amplitude.

In \cite{Romanelli2011NF} and \cite{Romanelli2013NF}, the flux-tube gyrokinetic simulations show that the core-peaked and high-charge-concentration Li$^{3+}$ contributes an outward impurity turbulent particle flux and a corresponding inward main-ion turbulent particle flux. The results in the global gyrokinetic simulations in this work are qualitatively consistent with those in \cite{Romanelli2011NF} and \cite{Romanelli2013NF}. The new finding in this work is that the oppositely directed impurity and main-ion particle fluxes can make the impurity TEF drive on ZFs even stronger than the main-ion TEF drive. This means that the impurity TEF drive cannot be ignored for the high-charge-concentration-lithium case. Note that equation \eqref{eq:Er gt} shows that the TEF drive carries the mass-to-charge ratio, $m_j/e_j$. The fully stripped light impurities, whose $m_{\rm z}/e_{\rm z}$ is close to $m_{\rm i}/e_{\rm i}$, are considered in this work, and the difference between the main-ion and impurity TEF drive is mainly determined by their magnitudes. However, $m_{\rm z}/e_{\rm z}>m_{\rm i}/e_{\rm i}$ is expected for partially stripped impurities (e.g., C$^{4+}$ and W$^{40+}$), and thus impurity TEF drive is predicted to be weighted by the larger mass-to-charge ratio for the partially stripped impurities.\\

\ack

This work was supported by the National Natural Science Foundation of China under Grant Nos, 12535014 and 12505261, and the Strategic Priority Research Program of the Chinese Academy of Sciences under Grant No. XDB0790201.

\appendix
\section{The detailed solution procedure of equation (1)}\label{appen:theory}
Following \cite{Wang2017POP} and \cite{RH1998PRL}, equation \eqref{eq:GKE} is solved by introducing $\partial_t\ll \mathcal {L}_{0,j}\sim 1/\tau_{b,j}$ for ZFs, with $\tau_{b,j}$ being the bounce time for trapped particles. Therefore, one can make the expansion $g_{j}=g_{j}^{(0)}+g_{j}^{(1)}$, with
\begin{align}
    \mathcal {L}_{0,j}g_{j}^{(0)}&=0, \label{eq:GKE-0ordering} \\
    \partial_tg_{j}^{(0)}+ \mathcal{L} _{0,j}g_{j}^{(1)}&=\frac{e_j}{T_j}F_{0,j}J_{0}\partial_t\delta\phi+J_{0}S_{j}F_{0,j}.\label{eq:GKE-1ordering}
\end{align}
Equation \eqref{eq:GKE-0ordering} indicates that $g_{j}^{(0)}$ should be a constant of unperturbed motion. $g_{j}^{(0)}$ is determined by the solvability condition, the orbital average of equation \eqref{eq:GKE-1ordering}\cite{Wang2017POP}. Assuming $g_{j}^{(0)}=g_{j,k}\exp({\rm i}k_rr)$, $g_{j,k}$ can be written as
\begin{equation}\label{eq:GKEsolution}
    g_{j,k}=e^{-{\rm i}\mathcal {Q}_j}\left[\frac{e_j}{T_j}\overline{\left(e^{{\rm i}\mathcal {Q}_j}J_0\delta\phi_k\right)}+\overline{\left(e^{{\rm i}\mathcal {Q}_j}J_0R_{j,k}\right)}\right]F_{0,j}.
\end{equation}
Here, $\mathcal {Q}_j=k_r\Delta r_j\equiv k_r(r-\overline{r})$, with $\overline{r}=\frac{1}{\tau_{b,j}}\oint {\rm d}\theta\frac{1}{\left(\textit{v}_{\parallel}\boldsymbol b+\bm{V} _{{\rm d},j}\right)\cdot\nabla\theta}r$. $R_{j,k}=\int {\rm d}tS_{j,k}$. Note that the distribution function for electrons is given by setting $J_0(k_r\rho_{{\scriptscriptstyle \rm L},{\rm e}})=1$ and $\mathcal {Q}_{\rm e}=0$, due to the much smaller Larmor radius and orbital width of electrons.\\

\section{A discussion of the response function $g(\tau)$}\label{appen:gt}
\begin{figure}[htbp]
\centering
\includegraphics[width=0.7\textwidth]{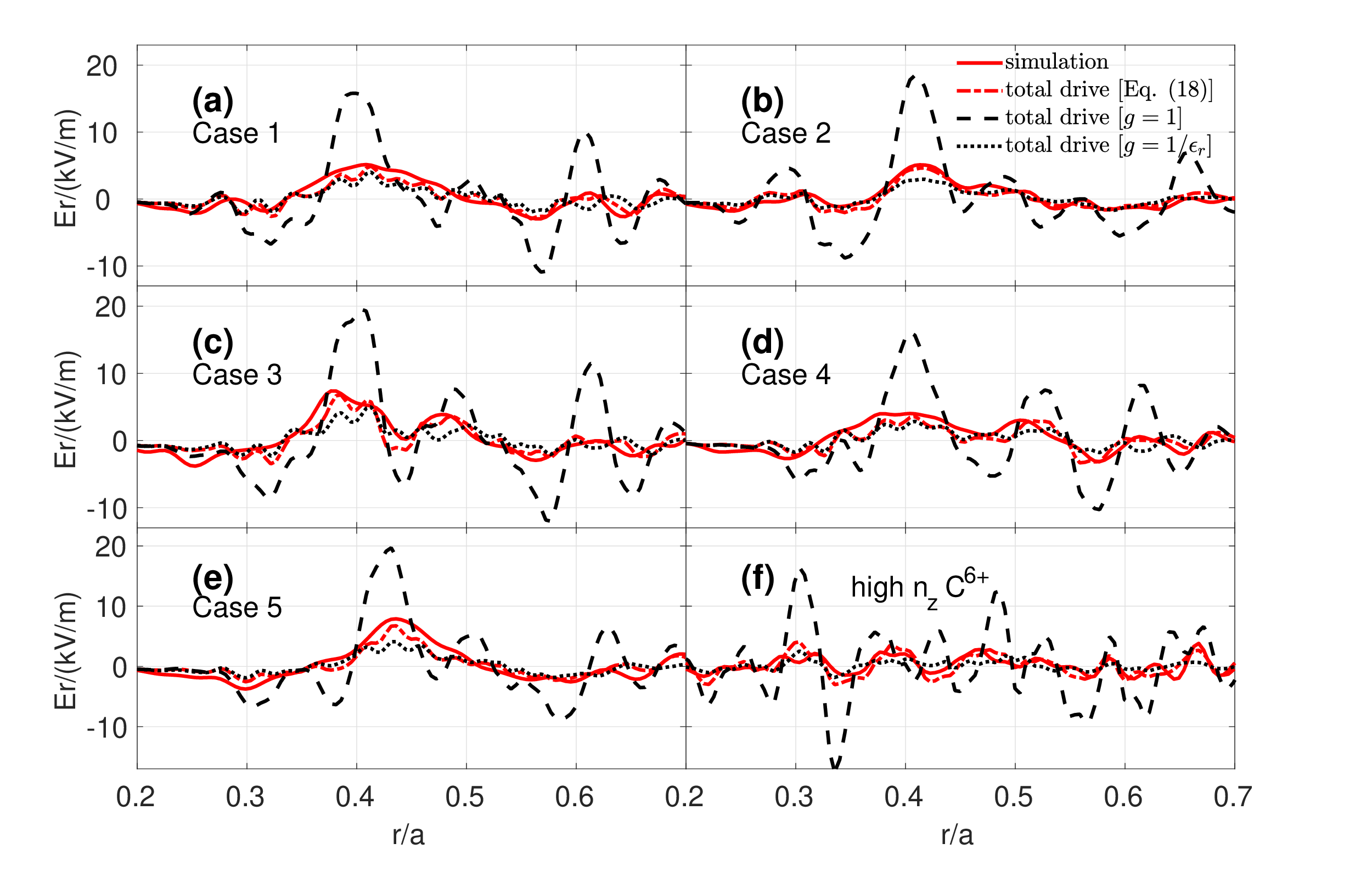}
\caption{Results of cases 1-5 [(a)-(e)] and an additional high-carbon-charge-concentration case (f) with different response functions $g(\tau)$. The simulation results are shown in red solid lines. The theoretical predictions obtained with equation \eqref{eq:gt} are shown in red dash-dotted lines, while those obtained with $g(\tau)=1$ and $g(\tau)=1/\epsilon_r$ are shown in black dashed and dotted lines, respectively.}
\label{fig:Er_shield}
\end{figure}

In this appendix, the theoretical predictions by equation \eqref{eq:Er gt} in cases 1-5 and an additional high-carbon-charge-concentration case [$\gamma_G/(C_{\rm s,i}/R_0)=0.17$] are presented in figure \ref{fig:Er_shield}. Three different response functions are compared, i.e., $g(\tau)=1$ (representing the turbulent PRS drive are unshielded), $g(\tau)=1/\epsilon_r$ (representing the turbulent PRS drive are totally shielded), and equation \eqref{eq:gt}. For cases 1-5 [figures \ref{fig:Er_shield}(a)-(e)], the theoretical predictions obtained with $g(\tau)=1$ (black dashed lines) and $g(\tau)=1/\epsilon_r$ (black dotted lines) do not agree with simulation results (red solid lines), while the predictions obtained with equation \eqref{eq:gt} (red dash-dotted lines) agree. Similar results were found in \cite{Wang2025NF}. The agreement between the equation \eqref{eq:gt} model and the simulation results indicates that the turbulent PRS drive is time-dependent, and the shielding timescale of the turbulent PRS drive is $1/\gamma_{G}$. In addition, for the high C$^{6+}$ charge concentration ($Z_{\rm z}n_{\rm z}/n_{\rm e}=0.86$) case with $\gamma_G/(C_{\rm s,i}/R_0)=0.17$, the theoretical prediction obtained with equation \eqref{eq:gt} also agrees with the simulation result [figure \ref{fig:Er_shield}(f)], which indicates that equation \eqref{eq:gt} also applies to the high impurity charge concentration case with a relatively lower GAM damping rate.

\bibliographystyle{unsrt}
\bibliography{reference}

\clearpage
\end{document}